\documentclass[twocolumn,showpacs,amsfonts]{revtex4-1}

\usepackage{amsmath}
\usepackage{latexsym}
\usepackage {mathrsfs}
\usepackage{float}
\usepackage{amssymb}
\usepackage{amsfonts}
\usepackage{graphicx}
\usepackage{textcomp}
\usepackage{hyperref}
\usepackage {mathrsfs}
\usepackage {pifont}
\usepackage {color}
 1

\newcommand{\dummy}

\begin{document}
\title{Effects of Domain Morphology on Kinetics of 
Fluid Phase Separation}
\author{Sutapa Roy and Subir K. Das$^*$}
\affiliation{Theoretical Sciences Unit, Jawaharlal 
Nehru Centre for Advanced Scientific Research, 
Jakkur P.O, Bangalore 560064, India}
\date{\today}

\begin{abstract}
Kinetics of phase separation in a three dimensional 
single-component Lennard-Jones fluid, that exhibits 
vapor-liquid transition, is studied via molecular 
dynamics simulations after quenching homogeneous 
systems, of different overall densities, inside the 
coexistence region. For densities close to the vapor 
branch of the coexistence curve, phase separation 
progresses via nucleation of liquid droplets and 
collisions among them. This is different from the 
evaporation-condensation mechanism proposed by 
Lifshitz and Slyozov, even though both lead to 
power-law growth of average domain size, as a 
function of time, with an exponent $\alpha=1/3$. 
Beyond a certain threshold value of the overall 
density, we observe elongated, percolating domain 
morphology which suddenly enhances the value of 
$\alpha$. These results are consistent with some 
existing theoretical expectations.
\end{abstract}
\maketitle

\section{Introduction}\label{introduction}
\par
\hspace{0.2cm}When a homogeneous system is quenched 
inside the coexistence curve the system phase 
separates into particle-rich and particle-poor 
domains. There has been longstanding interest in 
the kinetics of such processes \cite{binderbook,bray,
onuki,jones}. Important example systems are 
solutions of multi-component solids and fluids, 
vapor-liquid systems, etc. Depending upon the 
overall density ($\rho$) or composition (in case 
of a mixture), the phase separation can be 
spontaneous or the onset can get delayed. For 
quenches very close to the coexistence curve, the 
system is less saturated and so waits for droplet 
nucleation \cite{zettlemoyer,abraham,binder1,
kashchiev} events which occur via rare long 
wavelength fluctuations. The range of density or 
composition for which this is true is referred to 
as the nucleation regime. Beyond this the kinetics 
of phase separation is termed as spinodal 
decomposition \cite{onuki, jones} and the boundary 
between these two regimes is referred to as the 
spinodal line \cite{jones}, a concept drawn from 
mean field theory and may be valid for long 
molecular systems, e.g., polymers. Nevertheless, 
there have been attempts to draw this line even 
for small molecular systems.

\par
\hspace{0.2cm}During the phase separation, 
transport mechanism plays crucial role \cite{onuki} 
in the growth of average domain size ($\ell$). 
Typically, $\ell$ exhibits power-law 
enhancement with time ($t$) as \cite{binderbook,
bray,onuki}
\begin {eqnarray}\label{growthlaw}
\ell(t)\sim t^{\alpha}.
\end {eqnarray}
For diffusive transport, the growth exponent is 
$\alpha=1/3$, a value predicted by Lifshitz and 
Slyozov (LS) \cite{lifshitz,suman1,suman2}. For 
phase separation in solid mixtures, at moderately 
high temperatures ($T$), the LS value is the only 
exponent (at worst with minor corrections at small 
length limit) irrespective of whether the system 
is in nucleation or spinodal regime \cite{suman1,
suman2,suman3}. The situation in fluids, as 
described below, however, is not as simple. 

\par
\hspace{0.2cm}
Because of influence of hydrodynamics, the process, 
of course, is expected to be significantly faster 
in fluids. The hydrodynamic effects can show up 
either as an enhancement in the value of $\alpha$ 
or in the amplitude of growth. This fact is 
expected to be strongly dependent upon domain pattern.

\par
\hspace{0.2cm}For disconnected droplet morphology, 
close to the coexistence curve, Binder and Stauffer 
\cite{binder2,binder3} pointed out that the growth 
should occur via sticking of droplets following 
inelastic collisions. In that case, if the droplets 
exhibit random diffusive motion, the droplet density 
($n$) should decrease with time as \cite{siggia}
\begin {eqnarray}\label{BS1}
\frac{dn}{dt}=C_1D_\ell{\ell}{n^2},
\end{eqnarray} 
where $C_1$ is a constant and $D_\ell$ is the 
droplet diffusivity. Treating $D_\ell\ell$ as a 
constant, in accordance with the 
Stokes-Einstein-Sutherland (SES) relation 
\cite{hansen}, and using $n\propto \frac{1}{\ell^3}$, 
one obtains 
\begin {eqnarray}\label{BS2}
\frac{d\ell}{dt}\propto \frac{1}{\ell^2}.
\end{eqnarray} 
The solution of Eq. (\ref{BS2}) provides 
$\alpha=1/3$. Tanaka \cite{tanaka1,tanaka2,
tanaka3} argued that in high droplet density 
limit, motion of the droplets will not be random 
due to inter-droplet interaction. This was recently 
confirmed by us \cite{roy1,roy2}. Nevertheless, 
in this case also $\alpha$ should be $1/3$ but a 
difference will occur in the amplitude of growth. 

\par
\hspace{0.2cm}Note that the collision mechanism will 
be prominent once there are well formed droplets in 
the system. At early enough time, it is possible 
that the domains (with density or composition inside 
them reasonably away from the equilibrium coexistence 
value) are interconnected to give rise to a 
different mechanism and exponent. This we discuss next.

\par
\hspace{0.2cm}At high overall density or concentration 
of the minority component in a binary mixture, the 
interconnected morphology, mentioned above, remains 
there for all time. In this case, at the beginning, 
the simple particle diffusion mechanism, as in solids, 
plays the dominant role. With increasing domain size, 
of course, hydrodynamics is expected to take over. As 
suggested by Siggia \cite{siggia}, hydrodynamic 
mechanism, in this case, helps fast material flow 
through the tube like elongated domains due to the 
pressure gradient created by interfacial tension 
($\gamma$) in the undulated tube geometry. 
Consideration of Poiseuille flow in such a geometrical 
and physical situation will provide 
\begin {eqnarray}\label{viscous}
\ell=C_2t,
\end{eqnarray} 
where the constant of proportionality $C_2$ is a 
function of $\gamma$ and viscosity ($\eta$). Eq. 
(\ref{viscous}) is referred to as the viscous 
hydrodynamic growth law which at later time is 
expected to crossover to a slower growth with 
$\alpha=2/3$, referred to as the inertial 
hydrodynamic growth.
\par
\hspace{0.2cm}Eq. (\ref{viscous}) and other growth 
laws in domain coarsening systems can be obtained 
from simple dimensional arguments as well. Success 
of such approach is inherent in the fact that the 
systems, under discussion, remain invariant under 
appropriate time and length rescalings \cite{furukawa}.
\par
\hspace{0.2cm}Starting from the Navier-Stokes equation 
\cite{hansen}
\begin {eqnarray}\label{navier1}
\rho\frac{D}{Dt}\vec u-\rho\eta\nabla^2\vec u
=-\vec \nabla P,
\end{eqnarray} 
with $D/Dt=d/dt+(\vec u.\vec \nabla)$, $\rho$ the 
mass density, $\eta$ the kinetic viscosity, $P$ 
the pressure and $u$ being the fluid velocity, 
from dimensional substitutions, Furukawa 
\cite{furukawa} arrived at the equation 
\begin {eqnarray}\label{navier2}
\frac{d\ell}{dt}+f\ell\Big(\frac{d\ell}{dt}\Big)^2=
\frac{t_0\gamma}{\eta\xi\rho},
\end{eqnarray} 
which is a balance between frictions and force 
due to surface tension. In Eq.(\ref{navier2}), 
$f$ is a friction coefficient and $t_0$ is a 
relaxation time related to the equilibrium correlation 
length $\xi$ that has the critical divergence 
\begin {eqnarray}\label{epsilon}
\xi \sim \epsilon^{-\nu};~\epsilon=|T-T_c|,
\end{eqnarray} 
$T_c$ being the critical temperature and $\nu$ a 
critical exponent. The quantities $\xi$ and $t_0$ 
are used to make $\ell$ and $t$ dimensionless in 
Eq. (\ref{navier2}). From the decay of the dynamic 
structure factor \cite{hansen}, one can relate the 
time constant $t_0$ with $\xi$ as
\begin {eqnarray}\label{navier3}
t_0 \sim \frac{\xi^2}{D_\xi},
\end{eqnarray} 
$D_\xi$ being the diffusivity of clusters of the 
size of $\xi$. Again, using the SES relation 
\cite{hansen}, one can write
\begin {eqnarray}\label{navier4}
\eta \sim \frac{1}{\xi D_\xi}.
\end{eqnarray} 
Furthermore, in space dimension ($d$) three, 
$\gamma$ has the critical singularity \cite{privman}
\begin {eqnarray}\label{navier5}
\gamma \sim \xi^{-2}.
\end{eqnarray} 
All our results in this paper are from $d=3$. 
Combining Eqs. (\ref{navier3}-\ref{navier5}), one 
obtains \cite{furukawa} the right hand side of 
Eq. (\ref{navier2}) to be $\mathcal O(1)$. Note that 
in deriving Eq. (\ref{navier2}), an important 
assumption made was that there exist unique time and 
length scales in the problem that give rise to unique 
velocity, i.e., 
\begin {eqnarray}\label{navier6}
u=\frac{d\ell}{dt}.
\end{eqnarray} 

\par
\hspace{0.2cm}In Eq. (\ref{navier2}), Furukawa 
\cite{furukawa} identified the first and second 
terms on the left hand side as dissipative 
friction and inertial friction, respectively. 
Neglecting the inertial term, one obtains $\alpha=1$ 
and considering only the inertial term one gets 
$\alpha=2/3$. The lowering of the exponent from $1$ 
to $2/3$ can be associated with turbulence in the system 
because of which break-up of well grown domains 
into smaller ones is possible. The turbulent character 
of the system at late time means that the Reynold's 
number is domain-size dependent. Physically, for 
big enough domains, viscosity may fail to hold the 
domains. Nevertheless, in the competition between 
growth and break-up, the system effectively continues 
to grow.

\par
\hspace{0.2cm}Our primary objective in this work is 
to test the above mentioned morphology dependence of 
the hydrodynamic growth mechanisms and thus the 
growth laws, via computer simulations. For the sake of 
convenience, we have chosen vapor-liquid system. Note 
that in a binary or multi-component fluid mixture, one 
needs to choose a high overall density to avoid 
interference between liquid-liquid and vapor-liquid 
transitions. Further, for studies close to the 
coexistence curve, one needs very large systems to 
obtain reasonable droplet statistics. This fact, in 
case of a fluid mixture, can cause severe computational 
difficulty, since the overall density of the system 
must remain fixed.
\par
\hspace{0.2cm}Our observations in the vapor-liquid 
systems, that we have studied, confirm the above 
theoretical expectations for the morphology 
dependence of the growth exponents. We provide 
supportive results that the growth exponent 
$\alpha=1/3$, observed for low enough density, is due 
to a different mechanism than the LS mechanism. 
In addition, we present important results for various 
functions related to the pattern formation in 
coarsening systems.

\par
\hspace{0.2cm}The rest of the paper is organized 
as follows. In Section II we discuss the model and 
methods. Results are presented in Section III. 
Finally, the paper is summarized in Section IV.

\section{Model and Methodologies}\label{model}
\par
\hspace{0.2cm}In our model \cite{suman4} system, 
particles, of equal size and mass ($m$), at 
continuous positions $\vec r_i$ and $\vec r_j$ in 
a cubic box (unless otherwise mentioned) of linear 
dimension $L$ (in units of particle diameter), 
interact via 
\begin {eqnarray}\label{LJ1}
U(r=|{{\vec r}_i}-{{\vec r}_j}|)=u(r)-u(r_c)-(r-r_c)
{\frac {du}{dr}}\Big|_{{r}=r_c},
\end{eqnarray}
where
\begin {eqnarray}\label{LJ2}
u(r_{ij})=4\varepsilon[(\sigma/r)^{12}-(\sigma/r)^{6}]
\end{eqnarray}
is the standard Lennard-Jones (LJ) potential with 
$\varepsilon$ and $\sigma$ being respectively the 
interaction strength and particle diameter. In this 
model, the cut-off radius $r_c$ was introduced for the 
sake of computational benefit. The last term in 
Eq. (\ref{LJ1}) ensures the continuity of both 
potential and force at $r_c=2.5\sigma$. The critical 
temperature and critical density ($\rho_c$) for the 
vapor-liquid transition, that this model exhibits, 
were estimated \cite{suman5} to be 
$k_BT_c \simeq 0.9\varepsilon$ ($k_B$ being the 
Boltzmann constant) and $\rho_c \simeq 0.3$ 
(note that the density $\rho$ is defined as 
$\frac{N\sigma^3}{V}$, $N$ being the number of 
particles in the system of volume $V$). For further 
discussion, we set $k_B$, $\varepsilon$, $\sigma$ 
and $m$ to unity. This fixes the unit of 
time $\tau=\sqrt{m\sigma^2/\varepsilon}$ to unity 
as well. 

\par
\hspace{0.2cm}We have used molecular dynamics (MD) 
\cite{allen,frenkel} simulations to study coarsening 
phenomena in this model. All our simulations were 
done in the NVT ensemble. Unless otherwise mentioned 
we have used Nos\'{e}-Hoover thermostat (NHT) 
\cite{frenkel,hoover} for the control of temperature. 
In a recent paper \cite{roy2} we have demonstrated 
that NHT serves good purpose of studying hydrodynamic 
phenomena in this system. In the Verlet velocity algorithm 
\cite{frenkel}, that we have employed to solve the 
dynamical equations, the time step $\Delta t$ for 
integration was set to $\Delta t=0.005\tau$. All of our 
results were obtained by applying periodic boundary 
conditions and averaging over multiple initial 
configurations. 

\par
\hspace{0.2cm}Before quenching, homogeneous initial 
configurations were prepared at desired densities, 
via MD runs at temperature far above $T_c$. These 
homogeneous systems were quenched to $T=0.6$. As we 
will see, the coexistence vapor and liquid densities 
at this temperature are $\rho_{_v}^{\rm eq}\simeq 0.007$ and 
$\rho_{_\ell}^{\rm eq} \simeq 0.8$, respectively, 
giving a value of the coexistence diameter 
$\rho_{_d}\simeq 0.4$. We obtained results by varying 
$\rho$ between $\rho_{_v}^{\rm eq}$ and $\rho_{_d}$.

\par
\hspace{0.2cm}For the convenience of calculation of 
various observables we have mapped the original 
continuum configurations to lattice systems via the 
following procedure \cite{roy1,roy2}. From their original 
positions, particles were moved to the nearest sites 
of a simple cubic lattice of regular spacing $\sigma$. 
Following this, a filled lattice site $i$ was assigned 
a spin value $S_i=+1$ and for an empty site $S_i=-1$. 
Further, to deal with pure domain morphology, we have 
eliminated the thermal noise, over length scale $\xi$, 
via a majority spin rule \cite{suman1,suman2}. The 
analysis with this system thus becomes analogous to 
low-temperature Ising model. 

\par
\hspace{0.2cm}From these mapped configurations, 
the two-point equal-time correlation function, $C(r,t)$, 
was calculated as \cite{bray}
\begin {eqnarray}\label{cor1}
C(r,t)=\langle {{S_i}{S_j}}\rangle-{\langle {S_i}\rangle}
{\langle {S_j}\rangle},~~r=|\vec i-\vec j|.
\end{eqnarray}
In the rest of the paper we will deal with $C(r,t)$ 
normalized to unity at $r=0$. The average domain size 
$\ell$ was obtained from the decay of $C(r,t)$ as 
\begin {eqnarray}\label{cor2}
C(r=\ell,t)=h,
\end{eqnarray}
by setting $h=0.25$. Note that for self-similar 
pattern dynamics, $C(r,t)$ exhibits the scaling form 
\begin {eqnarray}\label{cor3}
C(r,t) \equiv \tilde {C}(r/\ell(t)),
\end{eqnarray}
where $\tilde C(\tilde x,t)$ is a master function 
independent of time. We have also calculated the 
domain size distribution function $P(\ell_d,t)$ where 
$\ell_d$ is the distance between two successive 
domain boundaries in $x-$, $y-$ or $z-$ directions. 
From normalized $P(\ell_d,t)$, $\ell$ was estimated as 
\begin {eqnarray}\label{prob}
\ell=\int d\ell_d \ell_d P(\ell_d,t).
\end{eqnarray}
We have checked that the results obtained from 
Eqs. (\ref{cor2}) and (\ref{prob}) are same, apart 
from a proportionality factor. So, in the rest of 
the paper we will stick to $\ell$ from 
Eq. (\ref{cor2}) only. Here we mention that 
$P(\ell_d,t)$ exhibits the scaling form \cite{bray}
\begin {eqnarray}\label{prob2}
\ell(t) P(\ell_d) \equiv \tilde {P}(\ell_d/\ell(t)),
\end{eqnarray}
where $\tilde P(\tilde y)$ is a time independent 
function.

\begin{figure}[htb]
\centering
\includegraphics*[width=0.4\textwidth]{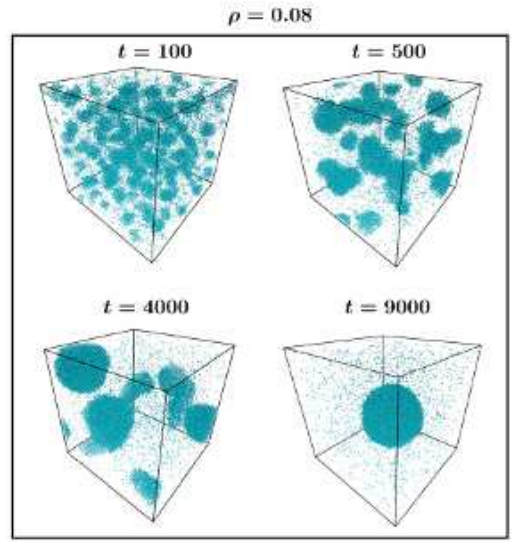}
\caption{\label{fig1}Evolution snapshots after 
quenching a homogeneous Lennard-Jones system, of 
density $\rho=0.08$, to $T=0.6$. The dots mark 
particle positions. Periodic boundary conditions were 
applied in all directions. The linear dimension of 
the cubic box is $L=80$.}
\end{figure}

\par
\hspace{0.2cm}In addition to the calculations 
discussed above, we have estimated various other 
quantities. Also important methodologies of analysis 
were used. These we will discuss in the next 
section while presenting the results.

\begin{figure}[htb]
\centering
\includegraphics*[width=0.4\textwidth]{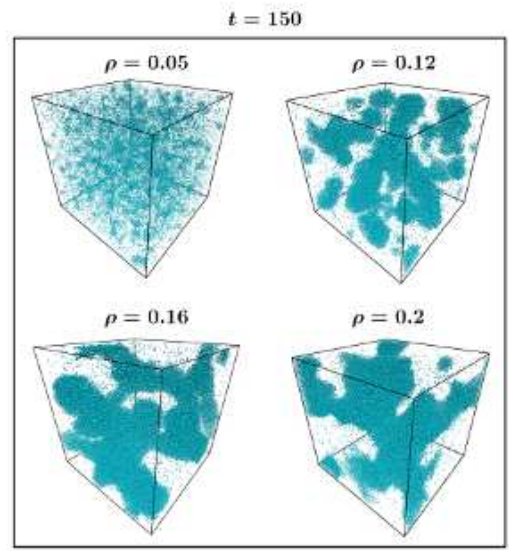}
\caption{\label{fig2}Snapshots during the 
vapor-liquid phase separation in the Lennard-Jones 
model with different overall densities. All results 
correspond to $t=150$ and $L=80$.}
\end{figure}

\section{Results}\label{results}
\par
\hspace{0.2cm}We begin with the presentation of 
evolution snapshots for an off-critical density 
$\rho=0.08$, in Fig. \ref{fig1}. Pictures are shown 
from four different times demonstrating the formation 
and growth of spherical liquid droplets in vapor 
background. Note that the dots in these snapshots 
represent particle locations. Here we point out that, 
due to fluctuations, the droplets have spherical 
symmetry only in average statistical sense. 

\par
\hspace{0.2cm}For an understanding of the density 
dependence of the domain pattern, in Fig. \ref{fig2} 
we show snapshots from four different densities, 
all from the same time. While for the lowest value of 
$\rho$ domains have just begun to form, for large 
enough $\rho$, the domains are quite robust. This 
indicates that the nucleation gets delayed with the 
approach towards the coexistence curve. 

\par
\hspace{0.2cm}From the snapshots of Fig. \ref{fig2}, 
it is clear that the densities, at least, upto 
$\rho=0.12$ are low enough so that the interconnected 
liquid domains are not stable. On the other hand, 
for $\rho \ge 0.16$, percolating domain structure is 
clearly visible. One may ask the question here, if 
the spinodal line should pass through a density 
between these values, at this temperature. Our 
primary objective in this paper, however, is to 
estimate the difference in domain growth mechanisms 
between the droplet and percolating morphologies. From 
Fig. \ref{fig2} we have identified a reasonable 
boundary between the two cases. Before moving onto the 
results and discussions for domain growth laws in the 
two cases, below we identify the coexistence 
densities at this temperature and check, based on the 
overall system density, during nonequilibrium evolution 
how fast the density in domains reaches those 
equilibrium values. In addition, we will also discuss 
results for few morphology characterizing functions. 

\begin{figure}[htb]
\centering
\includegraphics*[width=0.43\textwidth]{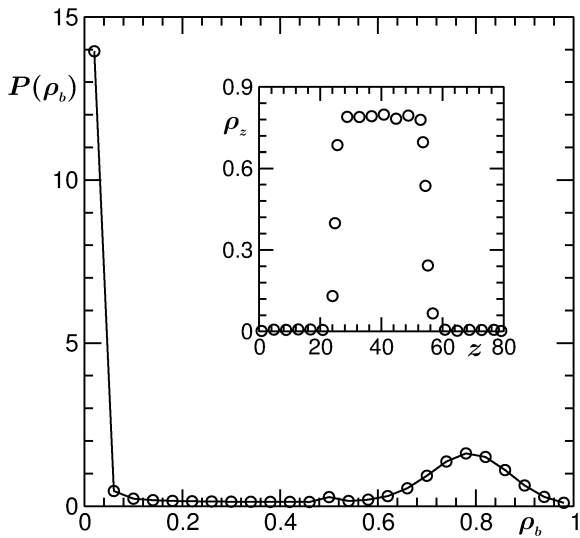}
\caption{\label{fig3}Plot of the probability 
distribution of density, $\rho_{_b}$, at different 
points in the system, for $\rho=0.3$ at $t=400$, 
with $L=64$. Note that the system is still far 
from equilibrium. Inset: Equilibrium density profile 
as a function of the $z-$ coordinate of an elongated 
box of $x-$, $y-$ and $z-$ dimensions $L_x=10$, 
$L_y=10$ and $L_z=80$, respectively. Again we have 
$\rho=0.3$. Note that apart from this one, all other 
results in the paper are from cubic boxes. The 
results are presented after averaging over $10$ 
initial configurations.}
\end{figure}

\par
\hspace{0.2cm}In the inset of Fig. \ref{fig3} we 
show the equilibrium density profile $\rho_{_z}$ 
as a function of $z$ coordinate of a box, elongated 
in this direction. Because of the shape of the box, 
from the energy minimization condition, it is 
obvious that in equilibrium the interface between 
the vapor and liquid phases will always be 
perpendicular to the $z-$ direction. From the flat 
regions of the plot we identify that 
$\rho_{_\ell}^{\rm eq} \simeq 0.8$ and 
$\rho_{_v}^{\rm eq} \simeq 0.007$. 

\par
\hspace{0.2cm}In the main frame of Fig. \ref{fig3} 
we show the distribution of local density, 
$\rho_{_b}$, obtained by constructing small boxes 
around a point including only its nearest neighbours. 
Note that this plot corresponds to nonequilibrium 
situation as opposed to the inset. Clearly two-peak 
structure is visible (due to small value of vapor-phase 
density, a maximum cannot be seen in the vapor side 
\textminus~for that one needs very small bin size 
which in turn will make the data very noisy). The 
nonzero width of these peaks are due to density 
fluctuations within the domains as well as because 
of contributions coming from the interfacial 
regions. One can, of course, obtain the average 
densities within the domains from the first moment 
of this distribution using data around each of the 
peaks. But the choice of the regions around the peaks 
is not completely unambiguous and also, the 
contribution from the interfacial regions cannot be 
separated out. So, we identify the location of the 
maximum as the average density (time dependent) within 
domains \textminus~this is reasonable considering the 
near symmetric look of the distribution, say, for the 
liquid part in Fig. \ref{fig3}. Note that both the 
main frame and the inset in this figure correspond 
to $\rho=0.3$. In the main part, the location of the 
liquid maximum is at $0.8$. This is consistent with the 
information obtained from the equilibrium picture in the 
inset which is due to the fact that the time chosen in the 
nonequilibrium case is rather late with respect to the 
pace of coarsening for this density \textminus~essentially 
by this time the domain order-parameter has equilibrated. 
But at early enough time there will be disagreement 
as seen below.

\begin{figure}[htb]
\centering
\includegraphics*[width=0.44\textwidth]{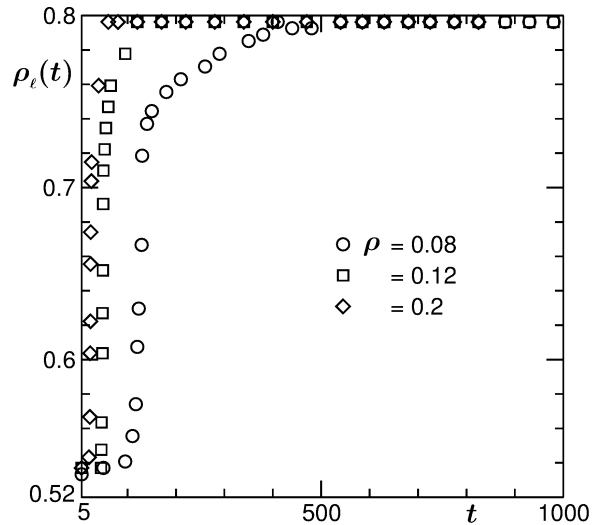}
\caption{\label{fig4}Plots of the average density, 
$\rho_\ell$, inside liquid domains vs $t$, for different 
values of $\rho$. All results correspond to an 
averaging over $10$ independent initial configurations 
with $L=80$.}
\end{figure}

\begin{figure}[htb]
\centering
\includegraphics*[width=0.44\textwidth]{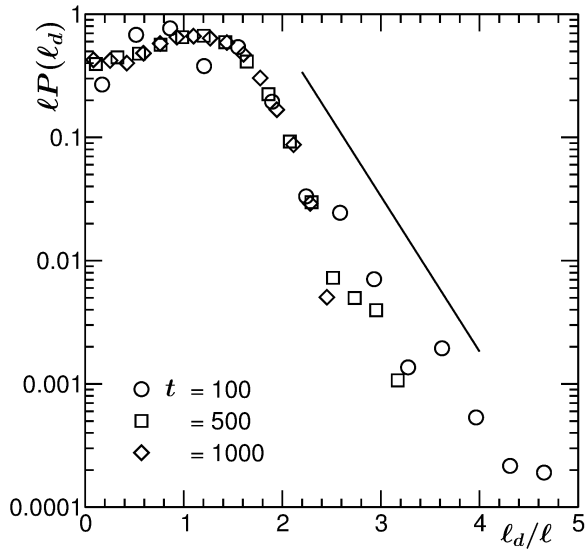}
\caption{\label{fig5}Log-linear plot of the master 
function, $\tilde P(\ell_d)$, vs the scaled 
ordinate $\ell_d/\ell$, for $\rho=0.08$. Data from 
three different times are shown. The solid line 
represents an exponential decay. The results 
correspond to $L=80$ and averaging over $10$ 
independent initial configurations.}
\end{figure}

\par
\hspace{0.2cm}In Fig. \ref{fig4} we present the liquid 
domain order parameter, $\rho_{_\ell}(t)$, as a 
function of time for different overall densities. 
It is clearly noticeable that $\rho_{_\ell}$ equilibrates 
faster with increasing supersaturation. Combining 
informations from Figs. \ref{fig2} and \ref{fig4}, we 
conclude that the time scales of equilibration within 
domain and that within system are smaller for higher 
densities. This, of course, is expected and will have 
consequence in the growth of $\ell$ at early time for 
small values of $\rho$. 

\par
\hspace{0.2cm}In Fig. \ref{fig5} we show the scaled 
plot of the domain-size distribution, on a log-linear 
scale, for $\rho=0.08$, using data from three different 
times. Reasonable collapse of data implies 
self-similarity of the domain structures at different 
times. The solid line in this plot corresponds to an 
exponential behavior. In the large domain-size 
limit $\tilde P$ is reasonably consistent with this 
\cite{das1,sicilia}. Flatter look of the data, for 
small abscissa variable, from later time is due to 
the fact that at very early time when the droplets are 
in the process of being nucleated, they have nearly 
uniform size and with the progress of time, the size 
dispersion increases giving rise to a larger width of 
the distribution.

\begin{figure}[htb]
\centering
\includegraphics*[width=0.44\textwidth]{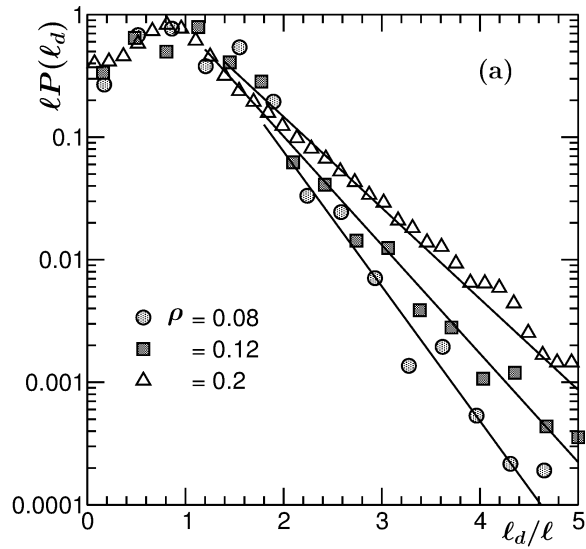}
\vskip 0.5cm
\includegraphics*[width=0.41\textwidth]{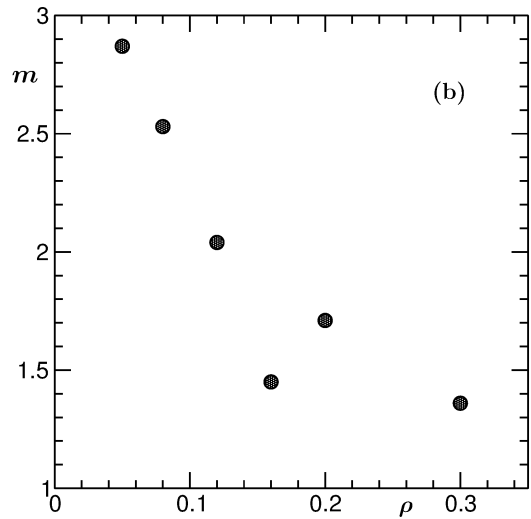}
\caption{\label{fig6}(a) Plots of the master 
domain-size distribution functions, $\tilde P(\ell_d)$, 
vs $\ell_d/\ell$ on log-linear scale, for three 
different values of density, at $t=100$. The solid 
lines correspond to fits to the tails of different 
data sets. (b) Plot of the slope, $m$, of the 
straight lines in (a), vs $\rho$. The system size 
and statistics are same as Fig. \ref{fig5}.}
\end{figure}

\par
\hspace{0.2cm}It will be interesting to compare 
the scaling function $\tilde P$ for various 
densities. In Fig. \ref{fig6}(a) we present results 
from three values of $\rho$. In all the cases 
exponential decay of the tails is visible, though 
they have different slopes on the log-linear plot 
\cite{das1}. In Fig. \ref{fig6}(b) we plot this 
slope $m$ as a function of $\rho$. It appears that 
$m$ sharply falls upto the percolation density and 
remains constant beyond that. In the droplet regime, 
this fall may be expected because, with increasing 
supersaturation larger droplets become available, 
implying larger dispersion.

\begin{figure}[htb]
\centering
\includegraphics*[width=0.44\textwidth]{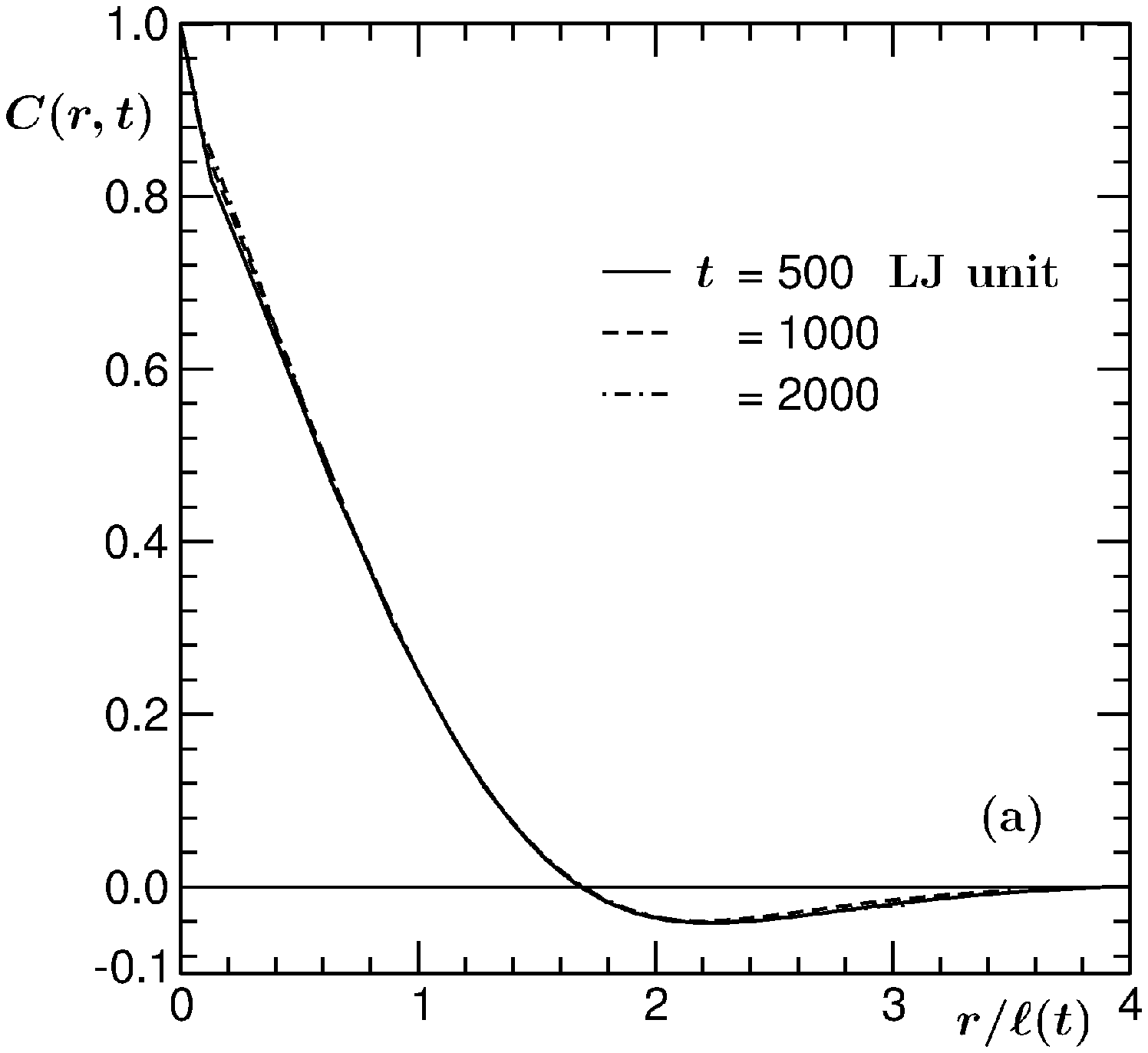}
\vskip 0.5cm
\includegraphics*[width=0.45\textwidth,height=0.46\textwidth]{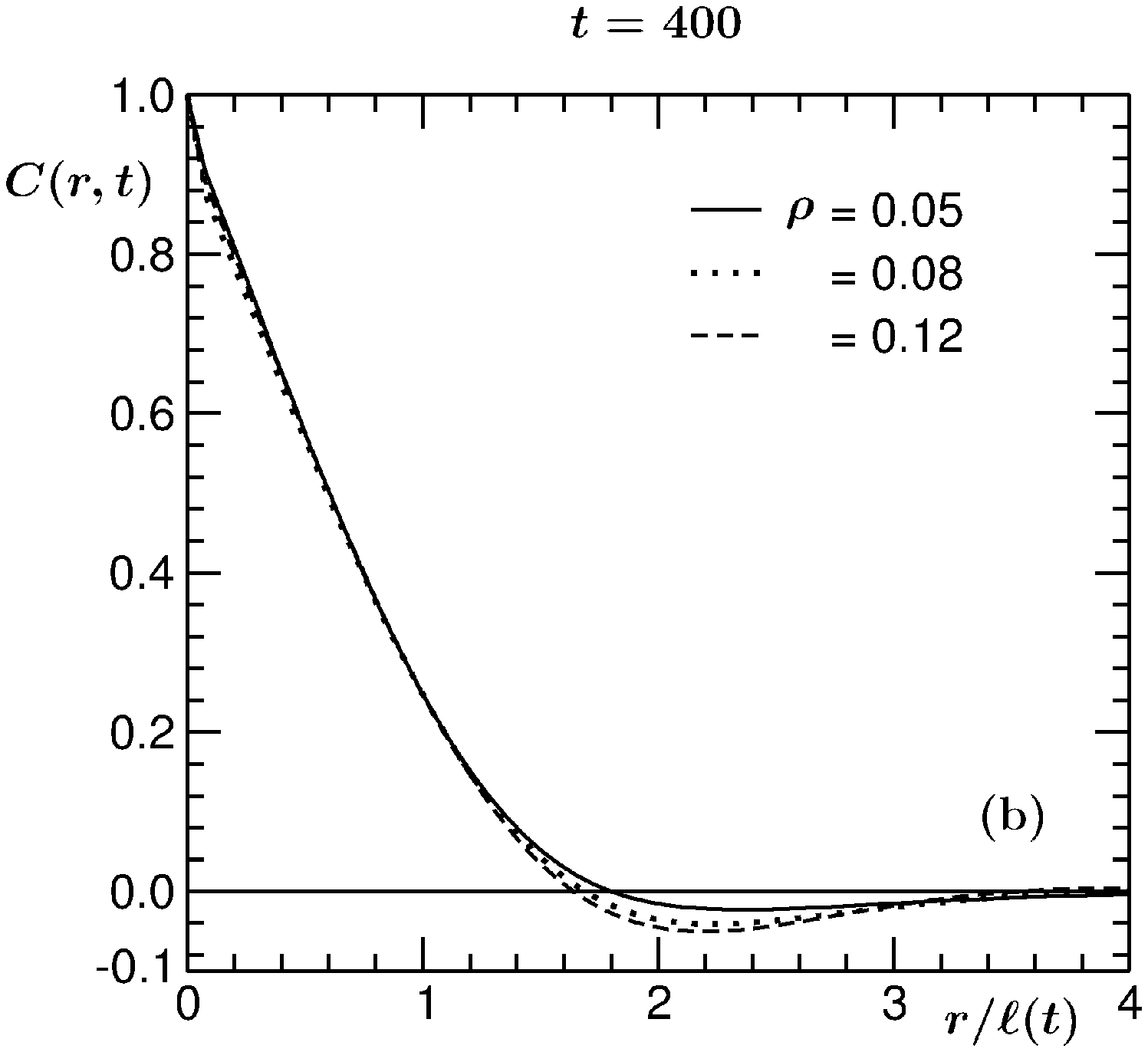}
\caption{\label{fig7}(a) Plots of $C(r,t)$, vs 
the scaled distance, $r/\ell(t)$, for $\rho=0.08$, 
from three different times. The values of $\ell(t)$ 
was obtained from the decay of $C(r,t)$ to $1/4$th its 
maximum value. (b) Plot of $C(r,t)$ vs $r/\ell(t)$, for 
different densities at time $t=400$. The results were 
obtained from an averaging over $10$ initial 
configurations with minimum value of $L=80$.}
\end{figure}

\par
\hspace{0.2cm}Exercises similar to Figs. \ref{fig5} 
and \ref{fig6} are done in Fig. \ref{fig7} but this 
time with the correlation functions. In Fig. \ref{fig7}(a) 
we show the plots of $C(r,t)$ as a function of the 
scaled distance $r/\ell(t)$ for $\rho=0.08$. Data from 
three different times are used, as indicated. Nice 
collapse of data again indicates self-similarity of 
patterns at different times. A comparative picture of 
this scaling function for different densities is 
demonstrated in Fig. \ref{fig7}(b). It is observed 
that the depth of the minimum increases with increasing 
density 
\cite{das1}.

\begin{figure}[htb]
\centering
\includegraphics*[width=0.43\textwidth]{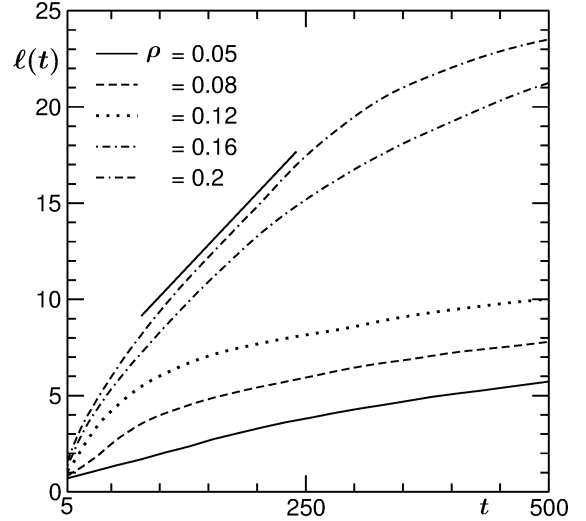}
\caption{\label{fig8}Plots of the average domain 
size, $\ell(t)$, vs $t$, on linear scale, for different 
densities. The solid line corresponds to a linear behavior. 
All the results are presented after averaging over 
$10$ initial configurations with lowest value of $L$ 
being $80$.}
\end{figure}

\begin{figure}[htb]
\centering
\includegraphics*[width=0.43\textwidth]{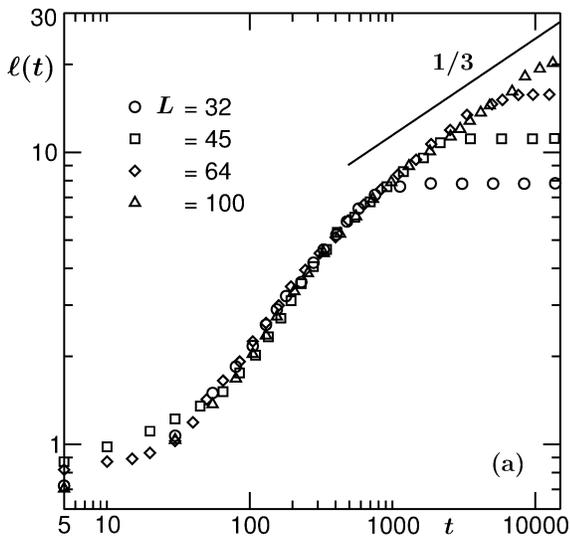}
\vskip 0.5cm
\includegraphics*[width=0.43\textwidth]{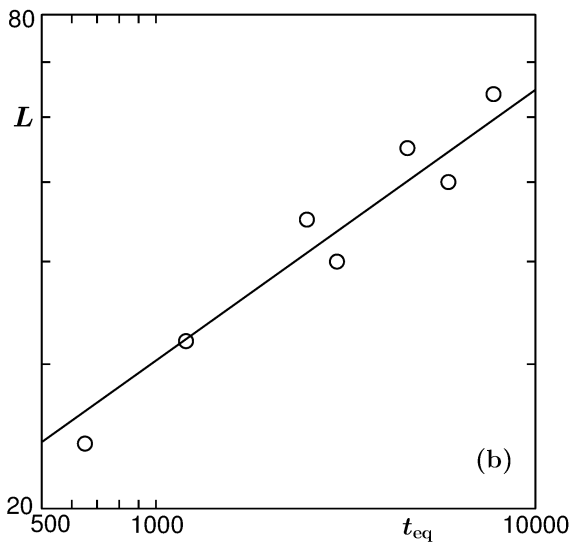}
\caption{\label{fig9}(a) Log-log plots of $\ell(t)$ 
vs $t$ for different system sizes and $\rho=0.05$. 
The solid line corresponds to a power-law growth with 
$\alpha=1/3$. (b) Log-log plot of system size, $L$, vs 
$t_{\rm eq}$ for $\rho=0.05$. See text for 
the definition of $t_{\rm eq}$. The continuous line 
corresponds to $\alpha=1/3$. In each of the cases the 
statistics of averaging is at least $6$ initial 
configurations.}
\end{figure}

\begin{figure}[htb]
\centering
\includegraphics*[width=0.47\textwidth]{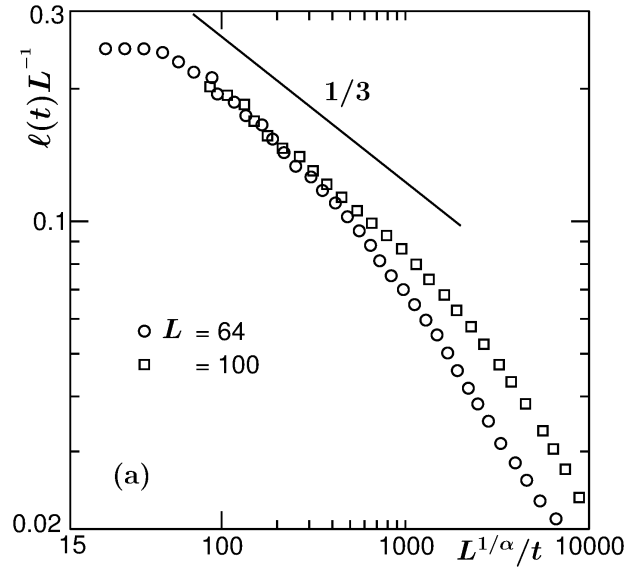}
\vskip 0.5cm
\includegraphics*[width=0.44\textwidth]{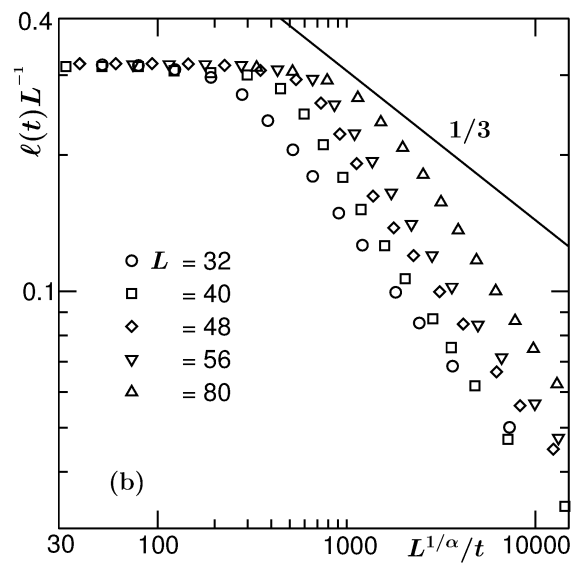}
\caption{\label{fig10}(a) Log-log plot of $Y(y)$ vs 
$y$ for different system sizes for $\rho=0.05$, by 
fixing $\alpha$ to $1/3$. The solid line corresponds 
to an exponent $1/3$. (b) Same as (a), but for 
$\rho=0.16$. The statistics in (b) is slightly poorer 
than in (a).}
\end{figure}

\begin{figure}[htb]
\centering
\includegraphics*[width=0.44\textwidth]{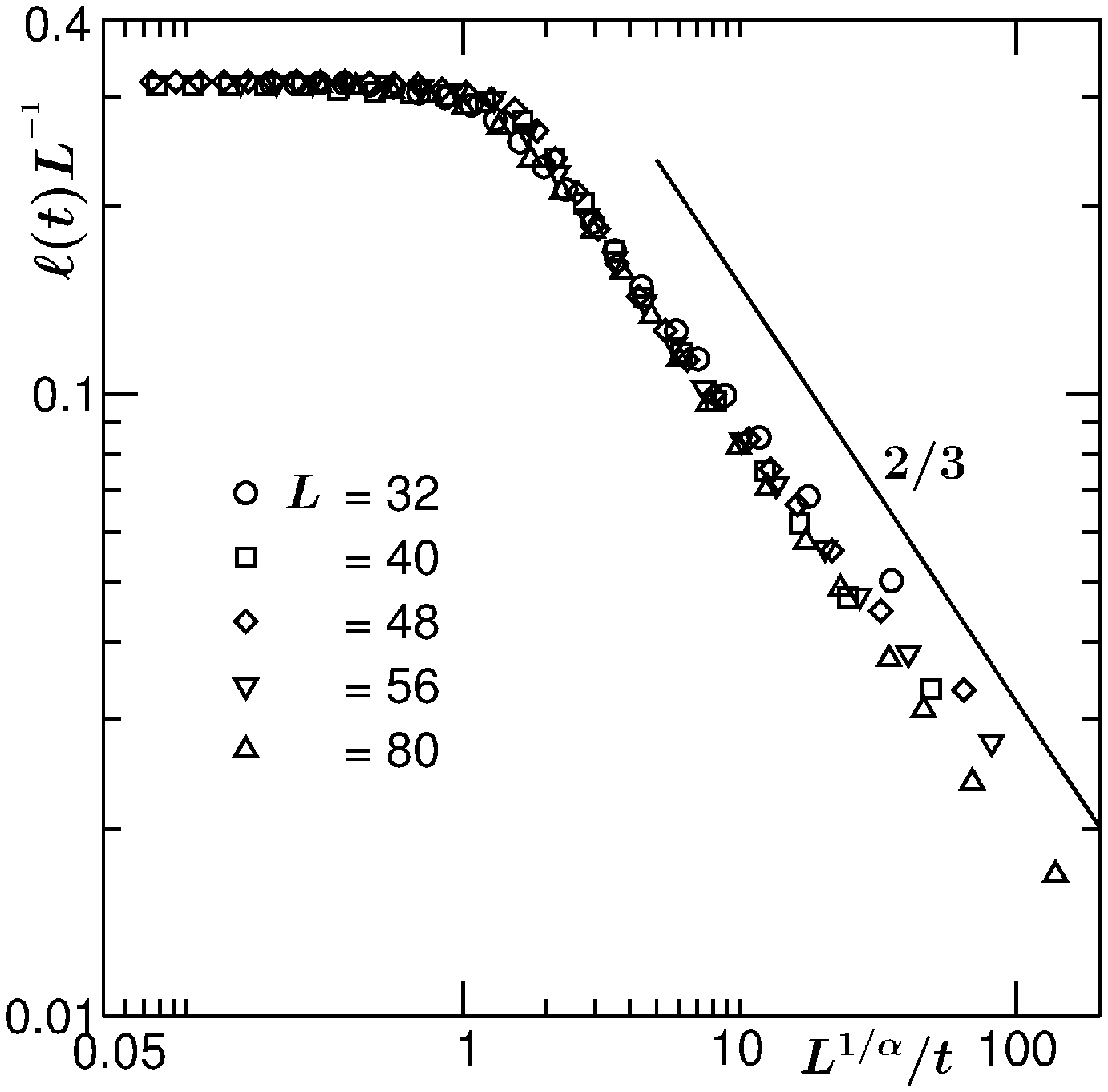}
\caption{\label{fig11}Log-log plot of $Y(y)$ vs $y$ 
for different system sizes for $\rho=0.16$, by fixing 
$\alpha$ to $2/3$. Here the solid line represents 
$\alpha=2/3$.}
\end{figure}

\par
\hspace{0.2cm}Next, in Fig. \ref{fig8} we show the 
plots of $\ell$ vs $t$ for few different densities. 
As is directly visible in Fig. \ref{fig2}, here also 
it is clearly seen that the growth occurs faster for 
higher density. Furthermore, there is a sudden jump in 
the growth rate between densities $0.12$ and $0.16$. 
As seen in Fig. \ref{fig2}, between these two values the 
morphological change from droplet to interconnected 
structure occurs. The consistency of the data set for 
$\rho=0.2$, over reasonable ranges of time and length, 
with the solid straight line is indicative of the fact 
that well above the percolation transition 
growth is linear in time. The deviation from this 
behavior at later time is due to the finite-size effects 
\cite{das2}. Here we note that the linear domain growth 
for the critical density in this model (for a slightly 
higher temperature) was confirmed via various different 
methods of analysis recently \cite{suman4,das2}. A 
nearly linear behavior is also seen in the data for 
$\rho=0.16$, however, we cannot confirm the actual 
exponent in this case from this plot. We will take a 
relook at it later. 

\begin{figure}[htb]
\centering
\includegraphics*[width=0.46\textwidth]{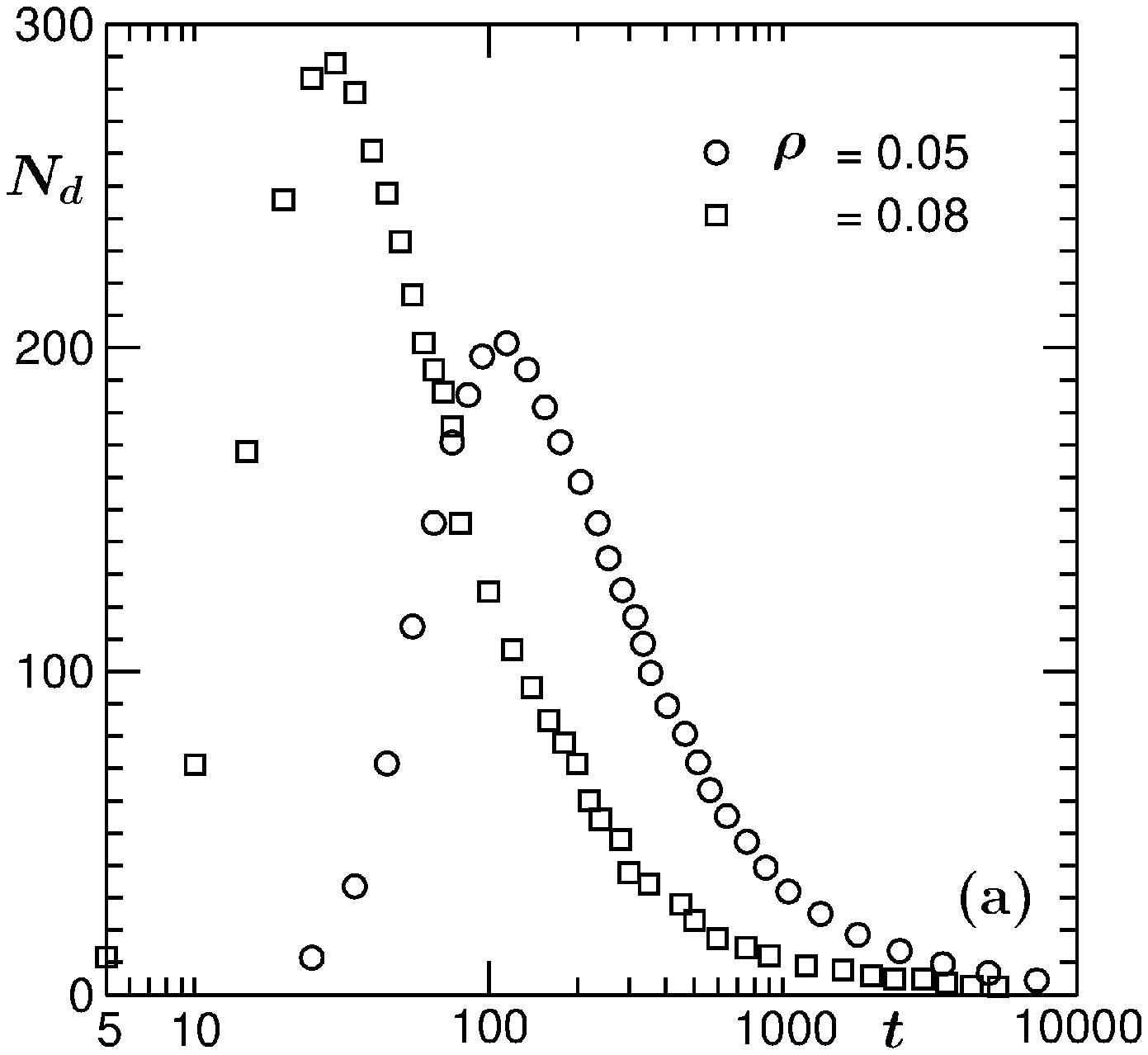}
\vskip 0.5cm
\includegraphics*[width=0.44\textwidth]{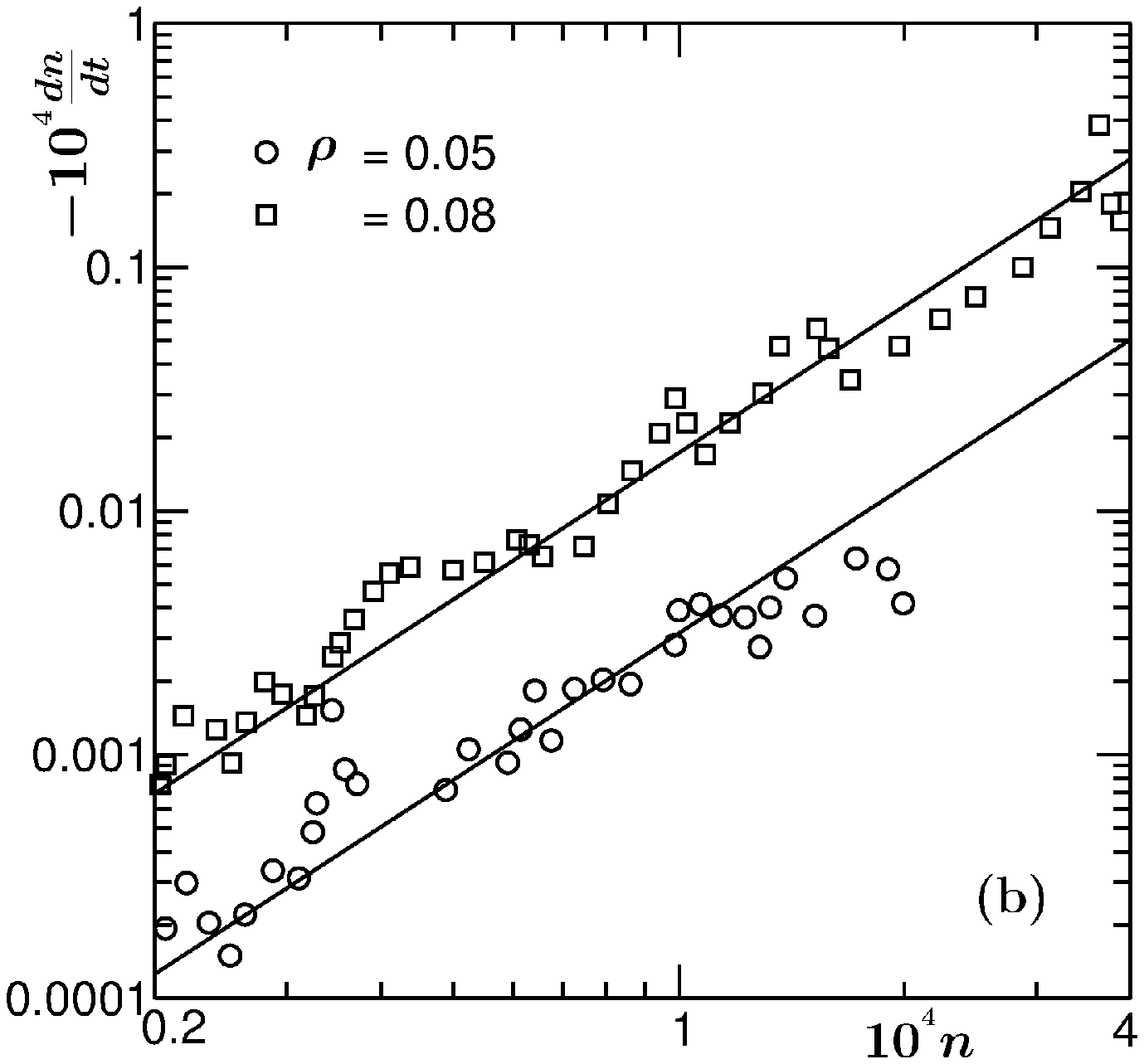}
\caption{\label{fig12}(a) Linear-log plot of droplet 
density, $n$, in the system, vs $t$, for two different 
values of $\rho$, as mentioned on the figure. (b) 
Log-log plot of $dn/dt$ vs $n$ for the same systems 
as in (a). The solid lines stand for quadratic ($n^2$) 
behavior. The ordinate for $\rho=0.08$ was multiplied 
by $5$.}
\end{figure}

\begin{figure}[htb]
\centering
\includegraphics*[width=0.44\textwidth]{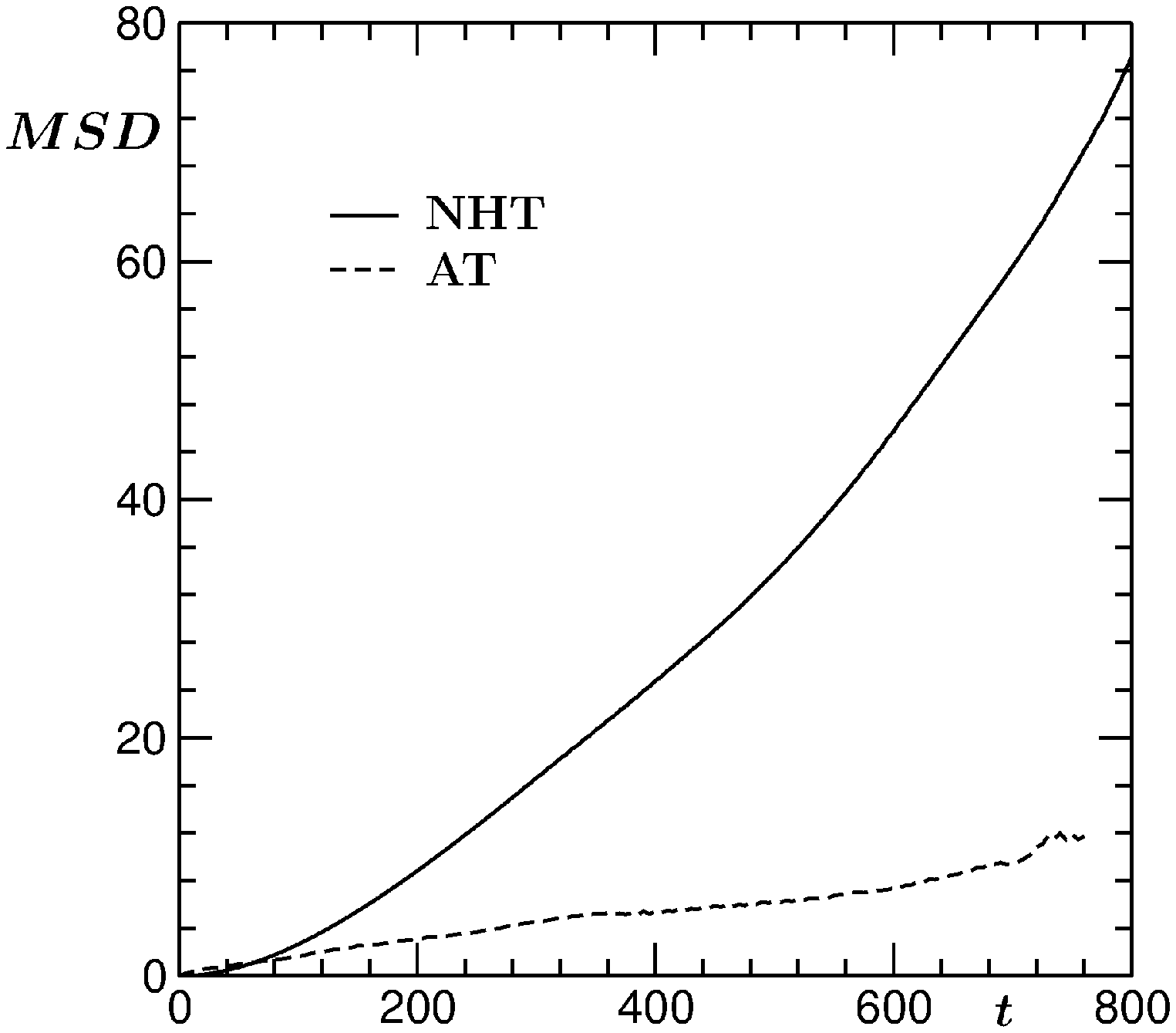}
\caption{\label{fig13}Plots of the MSD of droplets of 
approximately same size from NHT and AT, for 
$\rho=0.05$. Ordinate of data for AT has been 
multiplied by $50$.}
\end{figure}

\par
\hspace{0.2cm}In Fig. \ref{fig9}(a) we show the $\ell$ 
vs $t$ data for $\rho=0.05$ on a double-log plot. Results 
from various different system sizes are included. A 
flat part at the beginning indicates delayed nucleation. 
This is followed by a fast growth with exponent much 
higher than $1/3$. This can be attributed to the 
connectedness of unsaturated liquid domains at early 
time, as mentioned earlier. Finally, at very late time 
when there are well formed droplets the growth is consistent 
with an exponent $1/3$, over time range more than a decade. 
To further confirm the late time exponent value, we take 
the following finite-size scaling analysis \cite{fisher1} 
route. In this case, of course, the data in 
Fig. \ref{fig9}(a) clearly show an $1/3$ exponent. 
But such finite-size scaling exercise \cite{suman2}, 
as in Fig. \ref{fig9}(b), can prove useful in systems 
with strong finite-size effects. Here note that for a 
variety of phase separating systems, we have recently 
shown that the finite-size effects are rather weak 
\cite{das2}. Here also our results appear consistent 
with that picture. This can be appreciated from the fact 
that, in Fig. \ref{fig9}(a), almost all the way upto 
equilibrium, data for a smaller system follows that of 
a larger one.

\par
\hspace{0.2cm}In the analysis in Fig. \ref{fig9}(b), 
we use the fact that the equilibration time, $t_{\rm eq}$, 
for a system of size $L$ should scale with each other as 
\begin {eqnarray}\label{scaling1}
L \sim t^\alpha_{\rm eq}.
\end{eqnarray}
This is similar to finite-size scaling analysis of a 
quantity $X$ in equilibrium critical phenomena by 
computing it at finite-size critical temperatures 
and extracting the corresponding critical exponent from 
the plot of $X$ vs $L$. In the present case the inverse 
of equilibration time for a particular value of $L$ is 
analogous to the deviation of the finite-size 
critical temperature from the corresponding thermodynamic 
value. This latter quantity is related to $L$ via 
power-law with exponent $\nu$. We stress that the 
data in Fig. \ref{fig9}(b) reconfirms that the growth 
exponent is $1/3$. Similar exercises with other densities 
(with droplet morphology) also lead to the same value of 
the exponent. For the sake of brevity we do not show them 
here. Rather we reanalyze the same data set via another 
finite-size scaling method.

\par
\hspace{0.2cm}In equilibrium critical phenomena singularity 
of $X$ is quantified as 
\begin {eqnarray}\label{fss1}
X \simeq \xi^{x/\nu},
\end{eqnarray}
where $x$ is a critical exponent. Note that $\xi$, 
for a finite value of $L$, can at most be of the 
system size. So, at criticality 
\begin {eqnarray}\label{fss1.1}
X\sim L^{x/\nu}. 
\end{eqnarray}
Far away from $T_c$, one writes 
\cite{landau}
\begin {eqnarray}\label{fss3}
X=L^{x/\nu}Y(y);~y=\Big(\frac{L}{\xi}\Big)^{1/\nu},
\end{eqnarray}
where $Y$ is a scaling function independent of $L$. 
The behavior of $Y$ should be such that, for $T>>T_c$, 
one recovers Eq. (\ref{fss1}) involving the 
thermodynamic limit value of $\xi$, thus 
\begin {eqnarray}\label{fss4}
Y(y;y>>1)\sim y^{-x}.
\end{eqnarray}
In a plot of $XL^{-x/\nu}$, one uses $x$ as an 
adjustable parameter to obtain collapse of data from 
different sizes. The value of $x$ that provides the 
best collapse is taken as the thermodynamic value of 
the critical exponent.

\par
\hspace{0.2cm}In analogy with the above discussion of 
finite-size scaling in critical phenomena, one can 
construct similar equations for the present case as 
\cite{suman1,suman2,das2,heermann}
\begin {eqnarray}\label{fss5}
\ell(t)=LY(y),
\end{eqnarray}
\begin {eqnarray}\label{fss6}
y=\Big(\frac{L}{\ell}\Big)^{1/\alpha} \propto 
\frac{L^{1/\alpha}}{t}.
\end{eqnarray}
When $\ell/L$ is plotted vs $y$, only appropriate 
value of $\alpha$ will provide good collapse of data 
from different system sizes. The collapsed data in turn 
will have a behavior $\sim y^{-\alpha}$ in 
finite-size unaffected region. Here we emphasize that 
such finite-size scaling often becomes necessary to 
avoid less reliable methods of extracting exponents 
from log-log plots or data fitting exercises
that may suffer from inaccuracy due to 
presence of non-zero offsets or strong fluctuations 
in the data.

\par
\hspace{0.2cm}In Fig. \ref{fig10}(a), we plot 
$\ell/L$ vs $L^{1/\alpha}/t$, using two data sets 
of Fig. \ref{fig9}(a). From the behavior of data in 
Fig. \ref{fig9}(a), it is clear that there are multiple 
scaling regimes, as already stated. So, in this 
finite-size scaling exercise, we do not expect 
collapse of data in the whole range since only a single 
value of $\alpha$ can be used in this analysis. 
Since our objective for the droplet morphology is to 
establish the value $1/3$ at late time, we focus on 
obtaining optimum collapse in this 
regime only. Indeed, best collapse is achieved for 
$\alpha=1/3$ and the behavior of the master curve in the 
relevant region then is consistent with $y^{-{1/3}}$. 
In Fig. \ref{fig10}(b) we do the same exercise 
for $\rho=0.16$, again by fixing $\alpha$ to $1/3$. It is 
clear from this figure that $1/3$ is certainly not the 
right exponent for this density. 
\par
\hspace{0.2cm}
For $\rho=0.16$, desired behavior is best described 
for $\alpha=2/3$ which is shown in Fig. \ref{fig11}. 
This reconfirms that the exponents are very 
different for the droplet and percolating morphologies. 
However, a value different from $\alpha=1$, in this case, 
can be due to the fact that the connectedness of the 
morphology at this density, just above the threshold value, 
may not be very robust. So, the phase separation may 
progress via competition of growth and break-up of tubes. 
This, in fact, is the physical mechanism that leads to the 
inertial growth law. With the approach towards the 
coexistence diameter, these tubes become robust and the 
linear viscous hydrodynamic regime lives longer before 
crossing over to inertial regime. Nevertheless, here we 
only stress on the fact that the exponents in the 
droplet and percolating morphologies are certainly 
different and caution the reader that this value of 
$\alpha=2/3$ should not be taken very seriously. Note 
that at very early time there is a LS diffusive regime, 
however short it may be. In a more appropriate analysis 
the length and time of crossover from the diffusive 
regime to the hydrodynamic one need to be subtracted. 
However, because of the very low temperature chosen, it is 
difficult to identify these crossover parameters 
appropriately. If it becomes possible to do this analysis 
more accurately by incorporating this fact, one may 
obtain higher value of $\alpha$.

\par
\hspace{0.2cm}Even though, by now, we established that 
for droplet morphology $\alpha$ is $1/3$, this does not 
say that this value is the result of inter-droplet 
collision and not due to LS mechanism. In the following 
we proceed to resolve that. In Fig. \ref{fig12}(a) we 
present plots of droplet density $n$ as a function 
of $t$ for two different values of $\rho$. For a discussion 
on the identification of droplets and calculation of their 
number, we refer the readers to Ref. \cite{roy2}. This 
figure is quite instructive which again tells us 
that the nucleation gets delayed with decreasing overall 
density and clarifies why we have flat region in 
Fig. \ref{fig9}(a), at the beginning. The maximum in the 
plots is due to the fact that at early time nucleation 
events dominate the collision. 

\par
\hspace{0.2cm}From the plots of Fig. \ref{fig12}(a), 
we calculate $dn/dt$ and plot it vs $n$, on double-log 
scale, in Fig. \ref{fig12}(b). The data for both the 
densities are nicely consistent with power-law carrying 
exponent $2$. This confirms the validity of 
Eq. (\ref{BS1}) which was written down from the assumption 
of droplet collisions. 

\par
\hspace{0.2cm}Note that all our results so far were 
obtained by using NHT which preserves hydrodynamics well. 
Instead, if one applies a stochastic heat bath, e.g., 
Andersen thermostat (AT) \cite{frenkel}, it is expected 
that the growth will be in accordance with LS mechanism. 
In this latter case, we do not expect fast motion of the 
droplets. Rather, particles from smaller droplets will 
get deposited on larger droplets via diffusive motion, 
keeping essentially the droplet centre of mass (CM) fixed. 
In Fig. \ref{fig13} we have compared the mean squared 
displacements (MSD) of CMs of droplets of similar size for 
the NHT and AT cases. It is clearly seen that the above 
mentioned comparative picture is true. Note here that the 
AT data were multiplied by a factor ($>>1$). This 
difference between the two cases confirms our claim about 
the droplet motion and collision mechanism further.

\par
\hspace{0.2cm}Supersaturation in a vapor-liquid transition 
can be defined as \cite{wedekind} $S=\rho/\rho_{_v}^{\rm eq}$. 
Assuming that the spontaneous phase separation is 
related to the onset of interconnected structure, the 
value of $\rho$ should lie between $0.12$ and $0.16$. 
Taking $\rho=0.14$, we obtain $S\simeq 20$. Note that 
from an equation of state study of a similar model, 
the supersaturation at the spinodal point for slightly lower 
temperature (as a fraction of $T_c$ \cite{baidakov}) was 
obtained to be \cite{johnson} $S\simeq 32$. Our result can 
be treated consistent with that. Nevertheless, we again 
stress that the existence of a spinodal in a system 
like ours is of doubtful validity \cite{binder1}.

\section{Conclusions}\label{conclusion}
\par
\hspace{0.2cm}Via molecular dynamics simulations we have 
studied kinetics of phase separation for vapor-liquid 
transition, following temperature quenches inside the 
coexistence region for different overall densities. 
It is observed that for densities closer to the 
vapor branch of the coexistence curve, the late time 
growth dynamics is due to droplet motion and collision. 
On the other hand, for densities above a certain crossover 
value, domain morphology is percolating in nature and 
growth occurs due to fast motion of material 
through elongated tube like regions. These two mechanisms 
give rise to significantly different exponents, for the 
growth kinetics \cite{binder2,binder3,siggia,
tanaka1,tanaka2,tanaka3}, which are estimated.

\par
\hspace{0.2cm}Further, we have presented important 
results for the functions that characterize morphology. 
E.g., the domain size distribution exhibits exponential 
tail. It is demonstrated that the decay length of this 
function is strongly dependent upon the overall 
density. Similar results are presented for the 
two-point equal time correlation functions. Density 
dependence of these quantities are discussed.

\par
\hspace{0.2cm}Number of interesting further studies can 
be done. It will be important to look at the aging 
property \cite{fisher2,zannetti} in this system, 
particularly, its dependence on the variation of density. 
For the characterization of morphology, in addition to 
the scaling functions presented here, it will also be 
useful to calculate the fractal dimension \cite{strogatz}. 
We plan to address these questions in future. In addition, 
all these studies can be repeated for fluid mixtures as 
well \cite{lamorgese}, even though they will be 
computationally very demanding, as far as molecular 
dynamics simulations are concerned.

\vskip 1.0cm

\section*{Acknowledgement}\label{acknowledgement}
SKD and SR acknowledge financial support from the 
Department of Science and Technology, India, via 
Grant No SR/S2/RJN-$13/2009$. SR is grateful to the 
Council of Scientific and Industrial Research, India, 
for their research fellowship.
\vskip 0.5cm
\par
$*$~das@jncasr.ac.in


\vskip 0.5cm
\begin{thebibliography}{100}
\bibitem{binderbook}K. Binder, in \textit{Phase transformation of 
Materials}, edited by R.W. Cahn, P. Haasen and E.J. Kramer 
(VCH, Weinheim, 1991), Vol.\textbf{5}, p.405.
\bibitem{bray}A.J. Bray, Adv. Phys. \textbf{51}, 481 (2002).
\bibitem{onuki}A. Onuki, \textit{Phase Transition Dynamics} 
(Cambridge University Press, UK, 2002).
\bibitem{jones}R.A.L. Jones, \textit{Soft Condensed Matter} 
(Oxford University Press, Oxford, 2008).
\bibitem{zettlemoyer}\textit{Nucleation}, edited by A.C. 
Zettlemoyer (Dekker, New York, 1969).
\bibitem{abraham}F.F. Abraham, \textit{Homogeneous 
Nucleation Theory} (Academic, New York, 1974).
\bibitem{binder1}K. Binder, Rep. Prog. Phys. \textbf{50}, 
783 (1987).
\bibitem{kashchiev}D. Kashchiev, \textit{Nucleation: Basic 
theory with Applications} (Butterworth-Heinemann, Oxford, 2000).
\bibitem{lifshitz}I.M. Lifshitz and V.V. Slyozov, J. Phys. Chem. 
Solids \textbf{19}, 35 (1961).
\bibitem{suman1}S. Majumder and S.K. Das, Phys. Rev. E 
\textbf{81}, 050102 (2010).
\bibitem{suman2}S. Majumder and S.K. Das, Phys. Rev. E 
\textbf{84}, 021110 (2011).
\bibitem{suman3}S. Majumder and S.K. Das, in communication (2013).
\bibitem{binder2}K. Binder and D. Stauffer, Phys. Rev. Lett. 
\textbf{33}, 1006 (1974).
\bibitem{binder3}K. Binder, Phys. Rev. B \textbf{15}, 4425 (1977).
\bibitem{siggia}E.D. Siggia, Phys. Rev. A \textbf{20}, 595 (1979).
\bibitem{hansen}J.-P. Hansen and I.R. McDonald, \textit{Theory of 
Simple Liquids} (Academic Press, London, 2008).
\bibitem{tanaka1}H. Tanaka, J. Chem. Phys. \textbf{105}, 10099 (1996).
\bibitem{tanaka2}H. Tanaka, J. Chem. Phys. \textbf{107}, 3734 (1997).
\bibitem{tanaka3}H. Tanaka, J. Chem. Phys. \textbf{103} (6), 2361 (1995).
\bibitem {roy1} S. Roy and S.K. Das, Phys. Rev. E. \textbf {65}, 26141 (2002).
\bibitem{roy2}S. Roy and S.K. Das, Soft Matter, \textbf{9}, 4178 - 4187 (2013).
\bibitem{furukawa}H. Furukawa, Phys. Rev. A \textbf{36}, 2288 (1987).
\bibitem{privman}V. Privman, P.C. Hohenberg and 
A. Aharony, in \textit{Phase Transitions and Critical 
Phenomena}, edited by C. Domb and J.L. Lebowitz, vol. 14, Chap. 1, 
Academic Press, New York, 1991.
\bibitem{suman4}S. majumder and S.K. Das, Europhys. Lett. \textbf{95}, 
46002 (2011).
\bibitem{suman5}S. Majumder and S.K. Das, to be published.
\bibitem{allen}M.P. Allen and D.J. Tildesley, \textit{Computer Simulations 
of Liquids}, Clavendon, Oxford, 1987.
\bibitem{frenkel}D. Frenkel, B. Smit, \textit{Understanding Molecular 
Simulations: From Algorithm to Applications} (Academic Press, San Diego, 2002).
\bibitem{hoover}W.G. Hoover, in \textit{Studies in Modern 
Thermodynamics}, Elsevier, vol 11, 1991.
\bibitem{das1}S.K. Das and S. Puri, Phys. Rev. E \textbf{65}, 026141 (2002).
\bibitem{sicilia}A. Sicilia,Y. Sarrazin, J.J. Arenzon, A.J. Bray 
and L.F. Cugliandolo, Phys. Rev. E \textbf{80}, 031121 (2009).
\bibitem{das2}S.K. Das, S. Roy, S. Majumder and S. Ahmad, 
Europhys. Lett. \textbf{97}, 66006 (2012).
\bibitem{fisher1}M. E. Fisher, in \textit{Critical Phenomena}, 
ed. M.S. Green, Academic Press, London, 1971.
\bibitem{landau}D.P. Laudau and K. Binder, \textit{A Guide 
to Monte Carlo Simulations in Statistical Physics}, Cambridge 
University Press, Cambridge, 2009.
\bibitem{heermann}D.W. Heermann, L. Yixue and K. Binder, 
Physica A \textbf{230}, 132 (1996).
\bibitem{wedekind}J. Wedekind, G. Chkonia, J. Wolk, R. Strey and 
D. Reguera, J. Chem. Phys. \textbf{131}, 114506 (2009).
\bibitem{baidakov}V. G. Baidakov, S.P. protsenko, Z.R. Kozlova and 
G.G. Chernykh, J. Chem. Phys. \textbf{126}, 214505 (2007).
\bibitem{johnson}J.K. Johnson, J.A. Zollweg and K.E. Gubbins, 
Molecular Physics \textbf{78}, 591 (1993).
\bibitem{fisher2}D.S. Fisher and D.A. Huse, Phys. Rev. B \textbf{38}, 
373 91989).
\bibitem{zannetti}M. Zannetti, in \textit{Kinetics of Phase Transitions} 
(CRC Press, Boca Raton, 2009), edited by S. Puri and V. Wadhawan.
\bibitem{strogatz}S. H. Strogatz, \textit{Nonlinear Dynamics and Chaos: 
With Applications to Physics, Biology,Chemistry and Engineering}, 
Addison-Wesley, 1994.
\bibitem{lamorgese}A.G. lamorgese and R. Mauri, Physics of Fluids 
\textbf{17}, 034107 (2005).




\end{thebibliography}
\end{document}